\documentclass[]{interact}
\usepackage[dvipsnames]{xcolor}
\usepackage{epstopdf}
\usepackage{graphicx}
\usepackage{lscape}
\usepackage{rotating}
\usepackage{booktabs}
\usepackage{array}
\usepackage{pdflscape}
\usepackage{multirow}
\usepackage[table]{xcolor}
\usepackage{array}
\usepackage{booktabs}
\usepackage{comment}
\usepackage{natbib}
\bibpunct[, ]{(}{)}{;}{a}{}{,}
\renewcommand\bibfont{\fontsize{10}{12}\selectfont}
\theoremstyle{plain}

\theoremstyle{definition}

\theoremstyle{remark}

\usepackage{subcaption}
\usepackage{rotating}
\usepackage{pdflscape}
\usepackage{afterpage}

\begin{document}

\title{Dynamic production control and deadlock prevention in conveyor-equipped manufacturing systems}


\author{
\name{C. Castiglione\textsuperscript{a*}\thanks{*Corresponding author. Claudio Castiglione. Email: claudio.castiglione@polito.it}, E. Pastore\textsuperscript{a}, and A. Alfieri\textsuperscript{a}}
\affil{\textsuperscript{a}Department of Management and Production Engineering, Politecnico di Torino, Corso Duca degli Abruzzi, 24, Torino 10129, Torino, Italy}
}
\maketitle

\begin{abstract}
Flexible CONWIP approaches enable the dynamic production control to align production performance with production targets in the presence of high-fluctuating demand and system variability. However, frequently changing the WIP level might generate nervousness and, thus, can be practically infeasible. This paper investigates the combined use of job sequencing and routing to enable dynamic control in a CONWIP system comprising six unreliable workstations interconnected by conveyor carousels. Job sequencing and routing have been investigated using scenario analysis and discrete-event simulation. The results shed light on the effectiveness of the proposed approach, which improves throughput and enables dynamic production control without changing the WIP level. The results also provide managerial insights for system design, considering the impacts of the job-handling system (transport) and the minimum digital and technological requirements from Industry 4.0 for implementing the proposed approach.
\end{abstract}

\begin{keywords}
Production control; CONWIP; job sequencing; flow control; job-handling system; conveyor; 
\end{keywords}

\section{Introduction}
\label{sec:introduction}

The diffusion of the mass customization paradigm requires manufacturing systems capable of producing a large variety of product variants \citep{Hasani:2025} and, at the same time, reacting to more frequent fluctuations of product demands \citep{Naldi:2025}, energy prices \citep{Ostovari:2025}, critical raw materials price and lead times \citep{Castiglione:2025}. Manufacturing systems are no longer designed based on the requirements of a single product, but by considering at least a family of products \citep{Eswaran:2022}. Nonetheless, other products, neglected in the design phase, might be allocated to the system during its operational phase \citep{Friederich:2022}. 

The instability and unpredictability of customers' demand and production inputs, coupled with both the enhanced flexibility of manufacturing systems and their natural variability, jeopardize the effectiveness of production planning and control activities. These activities already aim to simultaneously improve and coordinate many different areas, including item production cycles, job-handling systems, buffers, and manufacturing resources \citep{Oluyisola:2022}. The current trends of mass customization \citep{Chorghe:2025} and global supply chain disruptions \citep{Bednarski:2025} exacerbate the challenges.

Production control policies, designed to maintain system performance in line with production targets \citep{Vespoli:2025}, must cope with such complexity while considering the underlying production paradigm (i.e., make-to-order or make-to-stock) \citep{Kim:2023}. All production control approaches, i.e.,  push, pull, and hybrid methods \citep{Geraghty:2005}, are adversely affected by rising complexity, primarily due to the amplification of variability propagation within the system \citep{Alfieri:2025}. However, the most affected methods are pull-based, as they leverage real-time information and local feedback mechanisms that become less reliable in more complex systems \citep{Jaegler:2018}.
   
Pull control strategies have evolved over the last decades to cope with fluctuations in production inputs, multi-product environments, and system variability \citep{Prakash:2015}. Among these, the approaches based on the constant WIP (CONWIP) control the WIP by keeping the number of jobs in the system constant. These approaches have been extended by implementing strategies to vary the number of cards to better adapt to external and internal changes: specifically,  adaptive \citep{Tardif:2001}, flexible \citep{Gupta:1997}, and reactive \citep{Takahashi:2004} strategies have been proposed. Also, CONWIP approaches have been combined with scheduling rules to improve effectiveness in more complex systems \citep{Thurer:2017}. 

Despite their ease of implementation, CONWIP approaches are not always the best performers in complex environments \citep{Jaegler:2018}. In fact, the multiple types of flexibility required in such systems increase the complexity of designing and implementing effective control policies \citep{Panzer:2022}. Moreover, the literature has highlighted the negative effect of frequently changing the CONWIP number of cards on variability propagation across interconnected systems \citep{Silva:2015}. Lastly, to the best of the authors' knowledge, there are few contributions in the literature that investigate CONWIP approaches in real systems, accounting for the interdependence between production and job-handling systems. 

This paper investigates, through Discrete Event Simulation and scenario analysis, the interdependent impacts of choices made at both the manufacturing and job-handling levels on system performance. In particular, it focuses on continuous-flow, fixed-path handling systems, such as conveyors, in a fully automated line controlled by a CONWIP policy. Defective products and quality inspection are also included in the analyses. Although the system operates in a make-to-stock environment, a CONWIP approach is adopted to mitigate disruptions caused by conveyor carousel congestion. The paper proposes a control mechanism to enhance the flexibility and responsiveness of the CONWIP policy in the presence of workstations interconnected by conveyors, with no buffers but carousels to enable job recirculation. A novel combination of the CONWIP policy with strategies of job routing and sequencing is proposed to avoid frequent fluctuations in the number of cards. Finally, this paper provides managerial insights for production planning and control, as well as for the design of fully automated production lines that jointly consider job-handling and manufacturing systems.

In the remainder of the paper, Section \ref{sec:literature} discusses the state of the art and Section \ref{sec:problemdescription} introduces the addressed problem. Sections \ref{sec:results} and \ref{sec:discussion} present the experimental results and their discussion, respectively. Section \ref{sec:conclusion} summarizes the concluding remarks.

\section{Literature Review}
\label{sec:literature}
The scientific literature has highlighted the poor performance of the standard CONWIP approach in dynamic environments \citep{Jodlbauer:2008}, identifying nine main critical aspects of CONWIP production control, which are introduced in the following. \cite{Hopp:2011} argued that in systems controlled by CONWIP approaches: (1) line order can be lost due to the cards movements through the workstations; (2) generally, backlog information is not collected nor exploited for WIP regulation; (3) determining the number of cards or the WIP level deserves depth analysis; there is a poor investigation of (4) shared resources, (5) multiple product families, and (6) non-linear process flows. \cite{Kim:2003} proposed a novel approach based on Dynamic Flow Control to overcome other two critical aspects of CONWIP approaches: (7) WIP levels can distribute within the system in an unbalanced fashion among the workstations, and (8) properly updating the control parameters of the CONWIP approaches (i.e., number of cards and loops) may require many internal (system) and external (demand) data. Last, \cite{Prakash:2015} argued that (9) the frequent onset of variability sources jeopardizes the effectiveness of continuous improvement actions, leading to a systematically under-performing system. 
However, these nine critical characteristics are closely dependent on other factors, such as the investigated system, the manufacturing paradigm in which it operates (e.g., make-to-stock or make-to-order), and the industrial context. 

This paper considers a single product make-to-stock system; thus, criticalities (1)-(2)-(5) do not apply as the manufacturing paradigm is make-to-stock and the jobs are considered identical. The paper focuses instead on (3)-(4)-(7)-(8)-(9), which are pivotal for this paper and all intertwined.  In fact, defining the number of CONWIP cards is complex, particularly due to the presence of non-linear and re-entrant flows, in which a single resource is shared among jobs at different production stages. As a consequence, the decision made at this level can unbalance the entire system, leading to changes in control parameters (e.g., the number of cards) and subsequent realignment due to the propagation of the variability related to machine reliability. Last, challenge (6) is partially addressed: the use of conveyor carousels leads to non-linear re-entrant flows, but no KPI related to delay (e.g., tardiness) is considered. 

Changes and fluctuations in product demand and system reliability, combined with variability propagation, affect the achievement of production targets and thus require parameter readjustment. In particular, most production control policies leverage the trade-off between WIP and resource capacity through approaches that are strictly dependent on the industrial field, the system itself, and the production paradigm in which it operates (e.g., make-to-stock or make-to-order) \citep{Prakash:2015}. Changes, fluctuations, and variability can affect the equilibrium among productive, protective, and excess WIP (and, similarly, available capacity), jeopardizing system throughput (TH) or causing congestion \citep{Blackstone:2002}. For these reasons, the dynamic adjustment of production control policies involves changing the WIP admitted in a specific area of the system: (i) directly by varying the number of cards, or (ii) indirectly by controlling job scheduling and routing. 

Direct approaches for controlling the amount of WIP across diverse parts of the system include CONWIP and Kanban-based approaches, which substantially modify the workload in each subpart of the system; a workload-based version of the CONWIP approach has been defined in \citep{Thurer:2019}. CONWIP approaches can also be combined with inventory and buffer levels, such as Basestock CONWIP \citep{Al:2018}. 

Indirect approaches primarily focus on control-flow mechanisms by identifying rules and methods for managing material flow through work centers \citep{Kim:2003b}. Control-flow mechanisms mainly consist of: (i) sequencing and scheduling of backlog, that is, the set of jobs waiting for the first available card that allows the system entry; (ii) sequencing and scheduling of jobs already flowing into the system; (iii) routing and transport rules across the workstations \citep{Lee:1997, Kim:2003b}. 

Since the advent of pull production control policies, the number of kanban cards is adjusted to address increases in system WIP levels or underutilization of production resources. \cite{Gupta:1997} proposed the first flexible kanban system to adapt the number of kanbans within the system to cope with stochastic processing times and demand fluctuations. \cite{Gupta:1998} introduced the dynamic kanban to minimize fluctuations caused by planned maintenance through adjusting kanban cards; \cite{Husseini:2006} adjusted kanban cards within the planning horizon through mathematical programming to enhance the just-in-time volume flexibility. 

Beyond the number of cards, control of an entire system can be achieved by partitioning it into subsystems with different CONWIP policies; thus, different ways to group workstations under CONWIP policies can alter workload allocation \citep{Huang:2016}. Moreover, subsystems can be simultaneously controlled using different pull approaches together, such as CONWIP and Kanban \citep{Ebner:2019}, or combinations of pull approaches with other control mechanisms, such as Drum-Buffer-Rope \citep{Kim:2003}.

In general, systems that specifically change the number of cards in CONWIP approaches were initially referred to: adaptive \citep{Tardif:2001}, flexible \citep{Gupta:1997}, and reactive \citep{Takahashi:2004}. Recently, flexible and dynamic CONWIP have been used interchangeably to refer to CONWIP approaches that adjust the number of cards over time, whereas the distinction between adaptive and reactive approaches hinges on the card-updating mechanism. Specifically, adaptive approaches adjust the card number in response to changes in certain parameters (e.g., demand), whereas reactive approaches define thresholds for parameters (e.g., demand, TH, WIP) that trigger the addition or removal of a predefined number of cards.

Table \ref{tab:litrev} summarizes the literature on approaches that enable adjusting system performance in response to changes in external or internal factors. In particular, the literature review examines the CONWIP control policy and its flexible variants (reactive and adaptive), also used in combination with scheduling or sequencing approaches applied to the backlog and to jobs already in the system. For all papers, the system configuration, the presence of re-entrant flows, the routing control mechanisms, and the job transport system are reported.

\afterpage{%
   \clearpage
\begin{table}[h!]
\caption{Dynamic approaches for controlling system performance within the CONWIP policy.}
\centering
\resizebox{0.7\textwidth}{0.9\textheight}
{%
\rotatebox{90}{%
\begin{tabular}{l|cccccccc}
\hline
\multicolumn{1}{c|}{\textbf{Reference}} &
  \textbf{System type} &
  \textbf{\begin{tabular}[c]{@{}c@{}}Re-entrant flows \\ (Y/N)\end{tabular}} &
  \textbf{\begin{tabular}[c]{@{}c@{}}Flexible Approach \\ (Ad/Re/No)\end{tabular}} &
  \textbf{\begin{tabular}[c]{@{}c@{}}Main control variable   \\ (WIP/Workload/Buffer)\end{tabular}} &
  \textbf{Transport} &
  \textbf{\begin{tabular}[c]{@{}c@{}}Scheduling approach\\ (Job/Backlog/No)-(Sequencing/Scheduling/No)\end{tabular}} &
  \textbf{\begin{tabular}[c]{@{}c@{}}Routing control \\ (Y/N)\end{tabular}} &
  \textbf{Objective function} \\ \hline
\cite{Bokor:2025} &
  Flow shop &
  N &
  Re &
  Workload &
  - &
  Capacity planning &
  N &
  Tardiness \\
\cite{Renna:2010} &
  Flow shop &
  N &
  Re &
  WIP &
  - &
  No &
  N &
  TH \\
\cite{Xanthopoulos:2019} &
  Flow shop &
  N &
  Re &
  WIP &
  - &
  No &
  N &
  TH/WIP \\
\cite{Belisario:2015} &
  Flow shop &
  N &
  Re &
  WIP &
  - &
  No &
  N &
  WIP/Backorders/Finished Stock \\
\cite{Hopp:1998} &
  Flow shop &
  N &
  Re &
  WIP &
  - &
  No &
  N &
  TH/CycleTime \\
\cite{Wurster:2025} &
  Flow shop &
  N &
  Re &
  WIP &
  - &
  Backlog-Scheduling &
  N &
  TH \\
\cite{Luh:2000} &
  Job shop &
  N &
  Re &
  WIP &
  - &
  Backlog-Scheduling &
  N &
  Lateness \\
\cite{E:2022} &
  Flow shop &
  N &
  Re &
  WIP &
  - &
  Backlog-Scheduling &
  N &
  TH/WIP \\ \hline
\cite{Ulhe:2024} &
  Flow shop &
  N &
  Ad &
  Workload &
  - &
  Backlog-Sequencing &
  N &
  TH \\
\cite{Azouz:2019} &
  Flow shop &
  N &
  Ad &
  WIP &
  - &
  No &
  N &
  Cost \\
\cite{Vespoli:2025} &
  Flow shop &
  N &
  Ad &
  WIP &
  - &
  No &
  N &
  TH/WIP \\
\cite{Gosavi:2024} &
  Job shop &
  N &
  Ad &
  WIP &
  - &
  No &
  N &
  Lead Time \\
\cite{Renna:2013} &
  Flow shop &
  N &
  Ad &
  WIP &
  - &
  No &
  N &
  TH/WIP/Flow Time \\
\cite{Liu:2009} &
  Flow shop &
  N &
  Ad &
  WIP &
  - &
  No &
  N &
  TH \\
\cite{Yoon:2013} &
  Flow shop &
  N &
  Ad &
  WIP &
  - &
  Backlog-Sequencing &
  N &
  Lateness \\ \hline
\cite{Thurer:2017} &
  Flow shop &
  N &
  No &
  Workload &
  - &
  \begin{tabular}[c]{@{}c@{}}Backlog-Scheduling\\ Job-Dispatching\end{tabular} &
  N &
  TH/Tardiness/Lateness \\
\cite{Dong:2016} &
  Flow shop &
  N &
  No &
  WIP &
  - &
  Job-Sequencing &
  N &
  Cost \\
\cite{Romagnoli:2015} &
  Flexible job shop &
  N &
  No &
  WIP &
  - &
  Backlog-Dispatching &
  N &
  Weighted Lateness \\
\cite{Framinan:2000} &
  Flow shop &
  N &
  No &
  WIP &
  - &
  Backlog-Sequencing &
  N &
  TH/CycleTime/Flow Time \\
\cite{Gailan:2023} &
  Flow shop &
  N &
  No &
  WIP/BUFFER &
  - &
  No &
  N &
  Cost \\
\cite{Al:2018} &
  Flow shop &
  N &
  No &
  WIP/BUFFER &
  - &
  No &
  N &
  Service Level \\
\cite{Ip:2002} &
  Flexible flow shop &
  N &
  No &
  WIP &
  Conveyor &
  No &
  N &
  TH \\
\cite{Yang:2011} &
  Flow shop &
  Y &
  No &
  WIP &
  \begin{tabular}[c]{@{}c@{}}Overhead Shuttle\\ Rail-guided vehicle\end{tabular} &
  No &
  N &
  TH/CycleTime \\ \hline
This paper &
  Flow shop &
  Y &
  No &
  WIP &
  Conveyor &
  Job-Sequencing &
  Y &
  TH \\ \hline
\end{tabular}%
}
\label{tab:litrev}
}
\end{table}
    \clearpage
}

The first set of papers in Table \ref{tab:litrev} leverages several techniques to develop reactivity into the CONWIP policy. Specifically, when the target system KPIs exceed or fall below fixed thresholds, the number of CONWIP cards changes accordingly. Reactive CONWIP is mainly applied to flow shop configurations in which all jobs have the same priority, and the main objective functions are related to TH, WIP, and cycle times. Exception are: \cite{Bokor:2025} that exploited the workload-oriented CONWIP policy in combination with the backlog earliest due date scheduling rule and a reactive approach for choosing the proper number of resources to minimize job tardiness; \cite{Luh:2000} that proposed a scheduling method for allocating jobs in a job shop by minimizing lateness within a CONWIP policy in which, for a given time window, WIP cannot exceed a maximum number of cards but it can be significantly less than the number of cards itself. Backlog scheduling was also exploited by \cite{Wurster:2025} in a job shop in which the system is controlled through a reactive CONWIP approach with a number of cards adjusted according to the planned capacity for managing order inventory, and by\cite{E:2022} that proposed a machine-learning-based approach for self-adapting WIP-cap to maximize system performance while reducing WIP and job average flow time given a set of orders waiting to be processed. All other reactive CONWIP approaches are implemented in a stand-alone fashion. \cite{Renna:2010} used a reactive approach to change the number of cards based on the current system utilization, while \cite{Xanthopoulos:2019} employed reinforcement learning to adjust the number of cards. \cite{Belisario:2015} proposed a combined use of discrete event simulation and genetic algorithm to derive a set of conditions to increase or decrease the number of CONWIP cards. Lastly, \cite{Hopp:1998} introduced a reactive CONWIP approach based on Statistical Throughput Control, in which periodic control over a sample of jobs updates the target TH and cycle time and adds or removes a card.

The second set of papers in the table includes adaptive CONWIP approaches, which are mainly focused on flow shop configurations and system performance rather than on the KPIs for individual jobs and orders. \cite{Yoon:2013} focused on job KPIs by combining an adaptive CONWIP approach for determining the number of cards based on the lateness of backlog with a backlog sequencing algorithm for minimizing the cycle times in a wafer fab. \cite{Ulhe:2024} leveraged backlog sequencing, specifically in a workload-oriented CONWIP policy, within a framework in which the workload is dynamically evaluated over time, identifying the optimal number of kanbans to maximize TH. On the one side, pure CONWIP adaptive approaches leveraged reinforcement mechanisms to vary the card number. \cite{Azouz:2019} proposed a neural network-based mechanism for adapting the number of cards to minimize performance cost while reducing the excessive number of fluctuations of card quantity; \cite{Vespoli:2025} exploited a deep reinforcement learning approach for controlling in an adaptive fashion the adjustment of card numbers to maximize TH while reducing system variability and WIP; and \cite{Gosavi:2024} developed an online machine learning algorithm that exploits closed formulation for identifying the number of CONWIP cards in order to reduce the nominal lead time of jobs within job shop systems. On the other side, the number of cards is adapted through analytical models: \cite{Renna:2013} proposed an adaptive CONWIP approach that adjusts the number of cards based on a controller that forecasts demand using two moving averages, one for the short-run forecast and the other for the medium-run; \cite{Liu:2009} studied an adaptive CONWIP approach based on the feedback control that compares the target TH with the actual TH to adjust the number of cards.

Finally, the last set of papers indirectly adjusts the workload across the system, thereby limiting the changes in the number of cards. In this set of papers, only \cite{Thurer:2017} focused on job KPIs by proposing a workload-based CONWIP approach that maximizes TH through the combined exploitation of backlog sequencing and job dispatching, while minimizing lateness and tardiness. Base-stock approaches have been used to improve dynamic control of the CONWIP policy. \cite{Al:2018} compared base-stock CONWIP and base-stock kanban CONWIP on an assembly line, where demand uncertainty is initially addressed through a genetic algorithm for line balancing. \cite{Gailan:2023} investigated (s, S), base-stock CONWIP, and base-stock policy considering low-estimated waste in perishable supply chain, showing that the base-stock CONWIP can improve the job average flow time by addressing the problem of system unbalancing within CONWIP control policy. Also, the material handling system has been used to dynamically control the system performance. \cite{Yang:2011} investigated the use of the CONWIP policy in a TFT-LCD manufacturing system characterized by re-entrant flows and served by rail-guided vehicles and overhead shuttles. The proposed static CONWIP approach initially identified the CONWIP loops; then, the number of cards for each loop was determined, considering transports and stockers, to minimize cycle time while improving TH. \cite{Ip:2002} adopted a CONWIP fixed structure in a flexible flow shop system with buffered workstations served by a conveyor, showing that higher processing time variability requires higher WIP to improve TH. The last important contribution to dynamic control of system performance without varying the number of cards concerns the scheduling and sequencing of jobs in the backlog and those already present in the system. In the literature, scheduling ans sequencing are implemented only in systems controlled by pure CONWIP policies: \cite{Dong:2016} combined CONWIP and a heuristic for internal job sequencing between the two stages in the shipbuilding sectors, demonstrating that properly managing workload within a CONWIP-controlled system can drive cost minimization; \cite{Romagnoli:2015} combined a customized dispatching rule for job release (backlog dispatching) with the CONWIP approach to maintain a constant WIP level while reducing lateness of high-priority jobs, and keeping a high utilization rate of the bottleneck resources in the multi-product make-to-order industry. \cite{Framinan:2000} showed that CONWIP policies, in a flow shop environment without re-entrant flows and transports, combined with different backlog sequencing approaches led to an improvement in system TH, both average WIP and flow time, in a flow shop environment. 

The recent literature on pull production control approaches has identified several critical aspects of the nervousness arising from frequent changes in the number of cards \citep{Azouz:2018}. In particular, dynamic CONWIP approaches can incur hidden costs due to frequent changes in the number of cards and may perform worse than expected in real environments \citep{Silva:2015}. In fact, varying the number of cards can be expensive (e.g., if cards are assigned to truck expeditions), or can involve unobserved materials (e.g., if cards are assigned to special supports or pallets) and resources (e.g., if cards are assigned to transport resources like AMR or AGV), or can propagate system variability to other systems (like warehouses) and misleading the management of costly resources (like inventories) \citep{Belisario:2015}.

\subsection{Contribution}
\label{sec:Contribution}
The literature review highlighted that very few contributions investigated how the combined investigation of job-handling systems and manufacturing lines affects CONWIP performance, particularly in the presence of re-entrant flows and rigid connections such as conveyors. This paper aims to make a first step in this direction. 

This paper investigates the effectiveness of a pure CONWIP strategy, simultaneously considering an automated manufacturing line and its job-handling system: a continuous-flow, fixed-path handling system that intertwines unbuffered workstations, where jobs waiting for the busy machine recirculate into automated conveyor carousels. 

The complexity of the investigated system and the role of the CONWIP policy are further underscored by the risk of deadlocks across the system carousels. Deadlocks are extreme congestion conditions that cause the entire automated line to stop, requiring time-intensive human intervention to restart production.

In this context, this paper investigates the combined use of flow control (routing) mechanisms and job sequencing to address the unbalancing caused by jobs recirculating among the carousels, and the variability introduced by defective products produced at each workstation, subsequently repaired and prepared for rework. 

Finally, this paper examines the effectiveness of integrating job-sequencing and system-handling-dependent flow-control mechanisms within a CONWIP policy to regulate system performance. Specifically, the proposed approach is assessed to determine whether the TH can be controlled without adjusting the quantity of cards by dynamically modulating production rates (i.e., either reducing or increasing performance) in response to system changes such as demand fluctuations, reduced machine reliability, the introduction of new products with distinct characteristics, elevated initial defect rates, or maintenance activities.

\section{Problem Description}
\label{sec:problemdescription}
The investigated production system is a simplified version of a real line for the production of stators for electric-vehicle engines, operating in a make-to-stock regime. \cite{Castiglione:2025} highlighted that the presence of conveyor carousels in this automated system exposes it to the risk of deadlocks due to both workstations and loop saturation. When the loop is congested, there may be no space left for additional jobs. In this case, when the machine finishes processing a job, there is no space on the conveyor to release it. Because the machine cannot release the job, it cannot pick up a new one, thereby reducing the queue (the number of jobs in the conveyor), and a deadlock occurs. When this happens, the line is blocked, and external intervention is required to restore system operationality \citep{Castiglione:2025b}.

Because of the CONWIP job-release mechanism, a new job can enter the system only when a completed (or scrapped) job leaves. In the analyzed system, jobs flow through the conveyor inside pallets; thus, every time a job leaves the system, the corresponding pallet is returned to the beginning of the line, where a new job can enter. 

The conveyor handling system is characterized by three carousels so as to create multiple loops among workstations. These loops play a crucial role, as they can serve as buffers for jobs awaiting processing on workstations. When a machine is busy, the job requesting it repeatedly circulates within the loop, rather than statically waiting in a buffer space, until the requested machine becomes available.

In this paper, the effectiveness of various job sequencing and routing policies for avoiding deadlock (and thereby increasing TH) and their dependence on the CONWIP parameter are evaluated. Job routing is related to the conveyor nodes, where product flow can be sent in two directions (i.e., split nodes). Split management policies can re-route some jobs from a congested loop to a different (even longer) route, thereby reducing congestion in that segment of the line. Additionally, sequencing rules at some workstations are implemented to alter how jobs are selected from the loop.  

\subsection{The physical system}
\begin{figure}[h]
\centering
\includegraphics[width=1\linewidth]{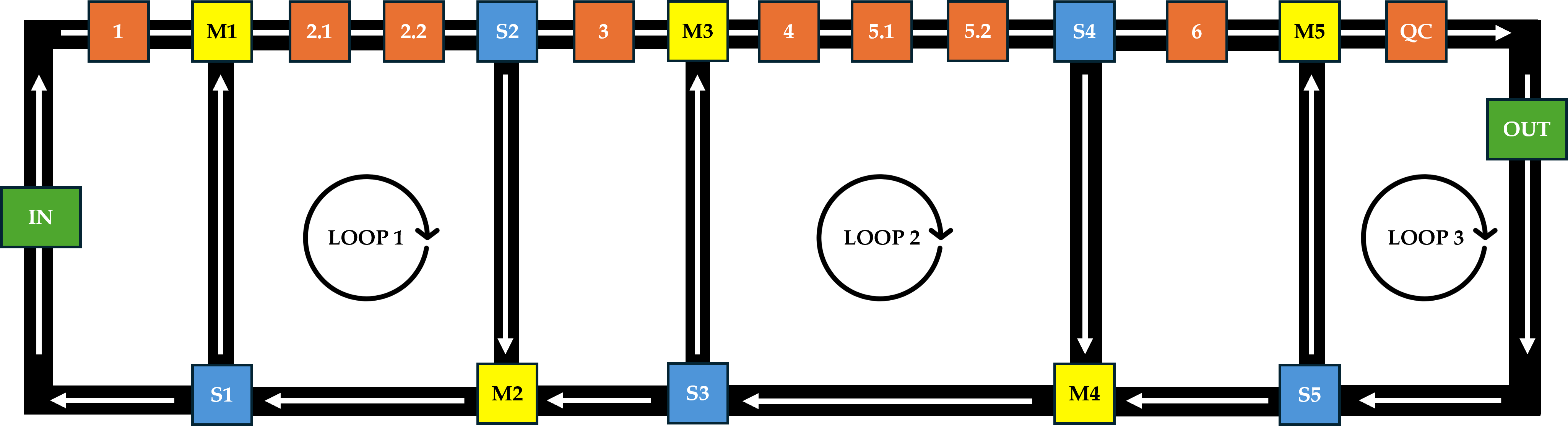}
\caption{Addressed assembly line}
\label{fig:prodlin}
\end{figure}

The line consists of six processing workstations (WSi, i = 1,..., 6) and a quality control workstation (WSQC) connected to each other by conveyor belts. Fig. \ref{fig:prodlin} shows the line graphical representation. Each of the six workstations (orange squares in the figure) performs a process step, while WSQC (orange QC square in the figure) performs quality inspection on jobs; jobs can be sent to WSQC after each process step. Two workstations, WS2 and WS5, have two identical machines (WS2.1-WS2.2 and WS5.1-WS5.2). There are two loading and unloading bays (green squares in the figure), i.e., areas where jobs enter the system to start their operations (IN) and exit at the end of their processes (OUT). Nodes S1 to S5 (light blue squares in the figure) represent splitters (points at which the jobs can take two different directions) while nodes M1 to M5 (yellow squares in the figure) represent mergers (points where two conveyor belts converge into a single flow). The figure shows conveyor belts in black and their directions indicated by white arrows. 

In the system in Fig. \ref{fig:prodlin}, the total number of pallets placed on the conveyor is constant and equal to the set CONWIP level, and jobs must be processed sequentially through the six workstations, from WS1 to WS6; WS5 requires two separate iterations per job. Whenever a new free pallet becomes available, a new job is placed on it, and the pallet begins flowing on the conveyor from the IN node to WS1. After being processed by WS1 and, when any WS2 machine becomes idle, the job can leave the workstation and move along the conveyor towards WS2 (i.e., WS1 has a blocking-after-service rule). The loop around WS2 (i.e., loop 1 in Fig. \ref{fig:prodlin}) is used to hold the jobs processed by WS2 until WS3 is available. After WS3, jobs enter loop 2 and circulate in it until WS4 is free. The same applies to jobs that have been processed in WS4/WS5 and are awaiting the first/second operation on WS5 (all jobs must be processed twice on either of the two machines in WS5). After the second operation in WS5, jobs are sent to WS6 when it becomes free, and in the meantime, they circulate in loop 2. Last, after being processed in WS6, jobs go to OUT, where they leave the system, while the related pallets return to IN onto the lower part of the conveyor.  

After being processed by each workstation except WS1, a sensor checks the processed job and, if any problem is detected, the job is sent to WSQC for an in-depth quality check; this can be considered a proxy for machine reliability and occurs with probability $\beta$. Jobs coming from all WSs circulate in loop 3 until WSQC is available. The inspection in WSQC can have three outcomes:
\begin{itemize}
    \item{rework:}
    the job is sent back to the origin WS (when a workable defect has been identified and removed);
    \item{discard:}
    the job is discarded as scrap (when a non-workable defect has been identified);
    \item{false positive:}
    the job is sent to the WS following the original one (when no defect has been identified) to proceed with the next operation.
\end{itemize}

In all the workstations (including WSQC), if the required resource is unavailable, the job continues circulating in the loop preceding or containing it (depending on the positions of splits and mergers), as previously described. Moreover, once a process is complete, the job can be repositioned on the conveyor belt to continue its journey if there is space available on the belt. If no space is available on the conveyor belt, the pallet with the job must remain on the machine, thereby blocking it; in this case, if the loop is already completely saturated, a deadlock occurs.



\subsection{Proposed control policies}
\label{sec:solutions}
Intuitively, the deadlock probability increases with the WIP circulating in the system. A simple solution to such a problem would be to reduce the CONWIP level (WIP*); however, as discussed in the literature, this would lead to significantly lower TH. Moreover, in a basic CONWIP environment, reacting to internal or external variability implies a change in the CONWIP level by changing the number of cards.

The objective of the control strategy proposed in this paper is to keep a fixed WIP* and to exploit control policies to either reduce the deadlock probability or to tune the TH to align with the short-run company target. In the following, split management policies and sequencing rules are proposed. How to efficiently combine them will be discussed with the numerical results in Section \ref{sec:results}. 


\subsubsection{Split management policies}
As congestion can occur across different loops, depending on working conditions, a way to reduce the occurrence of deadlocks without reducing WIP is to control the conveyor split nodes. By acting on splitters, it is possible to send jobs on different (and longer) paths, thus reducing the number of jobs in the already congested loops, i.e., in the loops when the deadlock probability is high.

In the addressed system, there are three loops (see Fig. \ref{fig:prodlin} as reference): 
\begin{itemize}
    \item Loop 1: it contains all jobs waiting to be processed at WS3;
    \item Loop 2: it contains jobs waiting to be processed at WS4, WS5 and WS6; 
    \item Loop 3: it contains only jobs waiting for the quality check at WSQC.
\end{itemize}
 Once a job enters a loop, it cannot return to previous loops. For example, jobs waiting to be processed at WS4 cannot exit loop 2 via S3 and return to loop 1. Thus, once downstream loops in the system (e.g., loop 3) are congested, there is no possibility of moving jobs back upstream to distribute them and thus avoiding deadlock. 

To overcome this problem, a split management rule is proposed for splitters S3 and S5, which are in loops 2 and 3, respectively. The idea behind the split management rule is that, when a loop is too congested (i.e., the WIP in that loop is high), the associated splitter opens, allowing jobs to move to upstream parts of the system until the loop is no longer congested. For this reason, no split policy is applied in loop 1, as 
the upstream part for it is only the system part containing workstation WS1, which is positioned on the conveyor and, hence, cannot be passed through by jobs already processed on it.

\begin{figure}[h]
\centering

\begin{subfigure}[t]{0.9\linewidth}
    \centering
    \includegraphics[width=\linewidth]{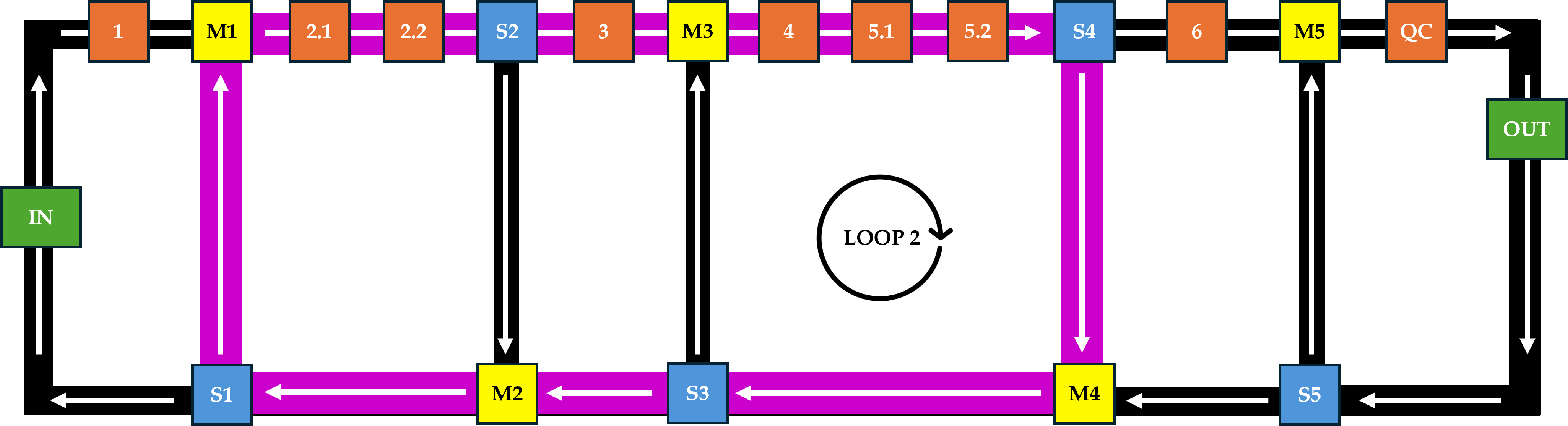}
    \caption{Job flow from opened loop 2 with S3 open.}
    \label{fig:loopB_expanded}
\end{subfigure}
\hfill
\begin{subfigure}[t]{0.9\linewidth}
    \centering
    \includegraphics[width=\linewidth]{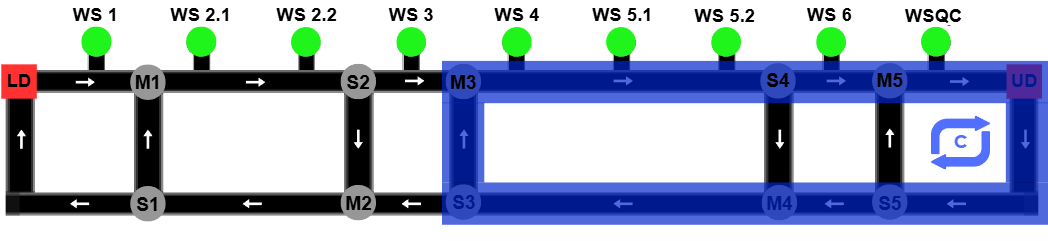}
    \caption{Job flow from opened loop 3 with S5 open (purple), and from opened loops 2 and 3 with both S3 and S5 open (pink).}
    \label{fig:loopC_expanded}
\end{subfigure}

\caption{Job flow changes when S3 and S5 are open.}
\label{fig:loops_expanded}
\end{figure}

Specifically, a split-opening threshold is fixed at loops 2 and 3, namely SO2 and SO3, controlling the behavior of splits S3 and S5, respectively. Every time a job circulating in loop $i$ waiting to be processed at the related WSs arrives at the related split (S3 for loop 2 and S5 for loop 3), the split checks whether the WIP at loop $i$ exceeds the SO$i$ threshold. If the condition holds, the split opens, and the job circulates back in the previous loop $i-1$; if the condition is false (i.e., the WIP in the loop does not exceed the threshold, so the loop is not congested), the job remains in the loop. Figures \ref{fig:loopB_expanded} and \ref{fig:loopC_expanded} show the new path jobs follow when S3 opens at loop 2 (Fig. \ref{fig:loopB_expanded}) and when S5 opens at loop 3 (Fig. \ref{fig:loopC_expanded}). The new job path can be different when S5 opens according to the split policy in loop 2: i) if S5 is open and S3 is closed, then jobs can circulate back from loop 3 to loop 2 (purple path in Fig. \ref{fig:loopB_expanded}); if S5 and S3 are both open, then jobs can circulate back to loop 1 (pink path in Fig. \ref{fig:loopB_expanded}). 


\subsubsection{Sequencing rules}
In the system, all workstations use a FIFO queuing rule to process jobs, and jobs move in the conveyor loops while waiting for the requested machine to be available.

Jobs are identical, and then can have the same priority across all WS queues, except for WS5 and WSQC. In WS5, jobs can be at either their first or second iteration, whereas in WSQC, jobs in the queue come from different origin WSs. 

To be coherent with the objective of acting towards the end of the system to distribute the congestion into the whole system (i.e., moving job congestion from the right of Fig. \ref{fig:prodlin} to a more balanced distribution through all the system), sequencing rules are only evaluated at the WSQC. Specifically, the FIFO queuing rule can be substituted by an Incremental (\textit{Inc}) sequencing rule. With \textit{Inc} sequencing rule, each job waiting to be processed at the WSQC (thus flowing in loop 3) and coming from WS$i$ is associated with the priority $p_i = \frac{i}{6}$.


\subsection{Design of Experiment}
\label{sec:DoE}
\begin{figure}[h]
\centering
\includegraphics[width=1\linewidth]{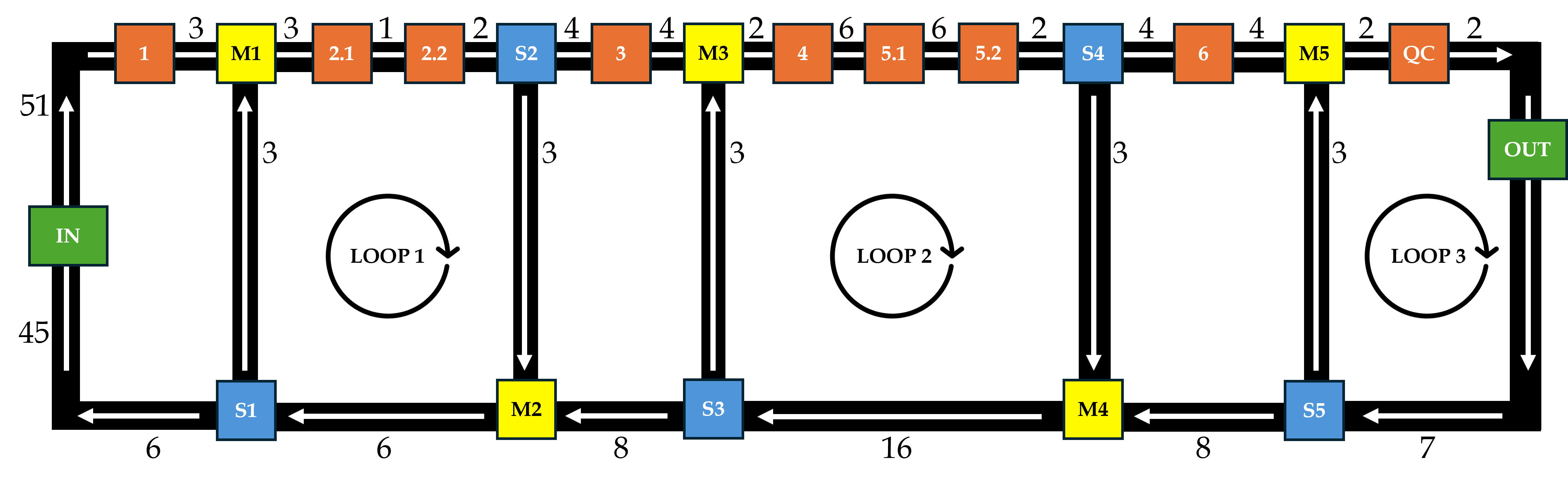}
\caption{Length of each segment of the network in units (each unit has the pallet size).}
\label{fig:conv_length}
\end{figure}

The lengths of the conveyor segments and their speeds were determined during the design phase of the line; hence, they are fixed. Fig. \ref{fig:conv_length} reports the length of each conveyor segment, in which the length unit refers to a single pallet dimension. The conveyor speed is set to 2 pallets per minute. 

To avoid confounding effects of parameter variability, while focusing on the effectiveness of job routing and sequencing rules combined with CONWIP policy, WS processing times are set to deterministic values, ensuring a balanced line. In each WS, the processing time is set to 20 minutes per job; to balance the line, each of the two identical machines in WS2 has a 40-minute processing time, and each of the two identical machines in WS5 has a 20-minute processing time (to account for the two necessary iterations per job). 

Regarding the quality check, the outcome is rework with probability 100\%. Instead, the $\beta$ probability for each job to go to WSQC after each WS (except WS1) is varied in the experiments.  

A Design of Experiment (DoE) has been developed to assess the impact of the proposed policies on system performance. The evaluated performance metrics are the average daily throughput (AVG dTH) and the deadlock (DL) probability.
The considered policies are: (i) the sequencing rule at the WSQC, (ii) the split management policy at loop 2, and (iii) the split management policy at loop 3. 

The DoE has five factors, namely: CONWIP level (WIP*), machine reliability (MR), WSQC sequencing rule (QCR), split opening threshold in loop 2 (SO2), and split opening threshold in loop 3 (SO3). 

\begin{table}
    \centering
    \resizebox{\textwidth}{!}{
    \begin{tabular}{cll}
         \textbf{Factor} && \textbf{Levels}  \\ \hline
         WIP* & CONWIP level & from 20 to 50 jobs, with an incremental step of 2 jobs   \\
         MR & machine reliability & 85\%, 88\%   \\
         QCR & WSQC sequencing rule & FIFO, Inc \\
         SO2 & split opening threshold in loop 2 & NO, 9, 19, 28, 37 \\
         SO3 & split opening threshold in loop 3 & NO, 3, 7, 10, 13 \\ \hline
    \end{tabular}}
    \caption{Design of Experiment: factors and levels}
    \label{tab:DoE}
\end{table}

Table \ref{tab:DoE} shows all the factors and their levels. \textit{WIP*} has 20 levels, ranging from 20 to 50 jobs, with an incremental step of 2 jobs; \textit{MR} has two levels, being 85\% and 88\%; \textit{QCR} can be either \textit{FIFO} or \textit{Inc}. Last, \textit{SO2} and \textit{SO3} take both numerical and \textit{NO} levels: the \textit{NO} level refers to no policy, i.e., the split at the related loop is not open; instead, the numerical values indicate that the split is able to open and the maximum WIP level at the related loop that enables the opening is equal to the numerical level value. About the combination of SO2 and SO3, the DoE does not include all the possible combinations of levels: specifically, the split in loop 2 can be opened only if the split in loop 3 is open. Thus, levels  9-19-28-37 for SO2 are combined only with numerical levels of SO3. In summary, the possible combinations of SO2 and SO3 are: both SO2 and SO3 equal to NO, numerical levels for SO2 and NO for SO3, and numerical levels for SO2 and SO3.

The total combinations of factor levels are 1344. For each combination, 100 simulation replicates are run, thus obtaining 134.400 experiments.

\section{Experimental results}
\label{sec:results}
\begin{figure}[h!]
\centering
\includegraphics[width=\linewidth]{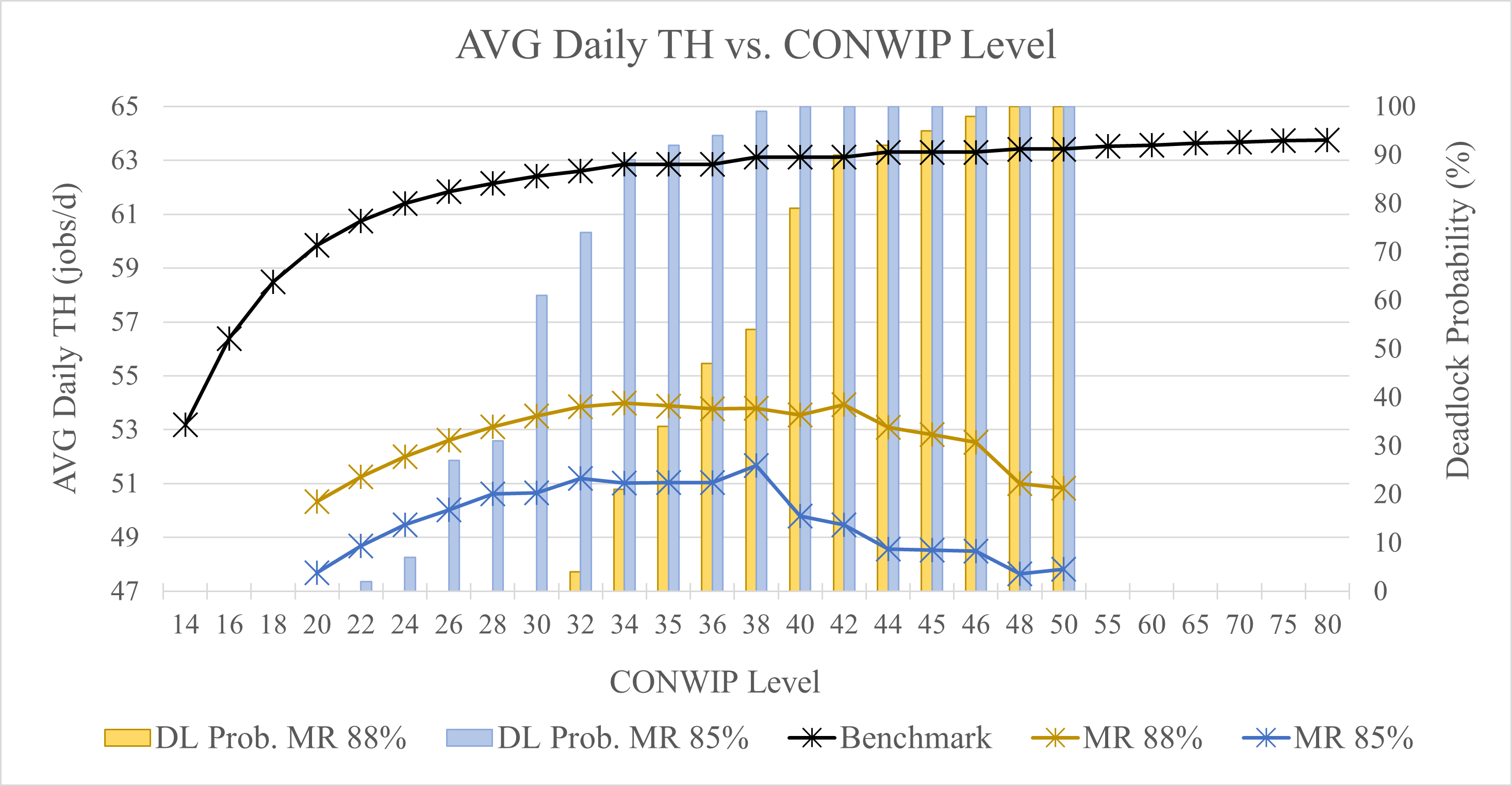}
\caption{Average daily throughput with different WIP* values for the three systems: benchmark, real system with machine reliability 88\% and 85\% (black, yellow, and blue lines, respectively). On the right, the secondary vertical axis shows the deadlock probability for systems with MR values of 88\% and 85\% (yellow and blue bars, respectively). Only scenarios with WSQC=FIFO and SO2=SO3=NO are considered.}
\label{fig:benchmark}
\end{figure}

Conveyor adoption drives and enhances system automation, but it also deeply affects line performance. Fig. \ref{fig:benchmark} shows, in black, the daily average TH for a benchmark version of the CONWIP system under investigation with different WIP* levels. The benchmark system has buffers with unlimited capacity, an MR level of 88\%, and part movements between workstations are modeled using a linear, continuous conveyor system without carousels. 

The real system employs conveyor carousels instead of buffers with unlimited capacity. In the figure, only scenarios with the FIFO sequencing rule at WSQC and no split management at both loop2 and loop3 are considered. For such cases, the AVG dTH is shown for systems with MR of 85\% and 88\% (blue and yellow lines in Fig. \ref{fig:benchmark}, respectively) for different WIP* values. Conveyor carousels negatively affect AVG dTH, reducing it by approximately 20\% (as shown by the difference between the benchmark black markers and MR 88\% yellow markers). 

A further effect of adopting conveyor carousels arises with the deterioration in machine reliability. The combined effect of carousels and a lower MR level results in an additional reduction in AVG dTH of approximately 5\% (as indicated by the difference between the blue and yellow lines in the figure). Moreover, it increases the probability of deadlock. The benchmark system has no deadlocks due to infinite-capacity buffers at each workstation; in the real system, with carousels and no buffers, deadlocks can occur when one or more carousels become congested. The deadlock probability is shown in Fig. \ref{fig:benchmark} with yellow and blue bars for MR=88\% and MR=85\%, respectively. The figure shows that a lower MR value increases the deadlock probability, as does a lower WIP*. 

\begin{figure}[b!]
\centering
\includegraphics[width=\linewidth]{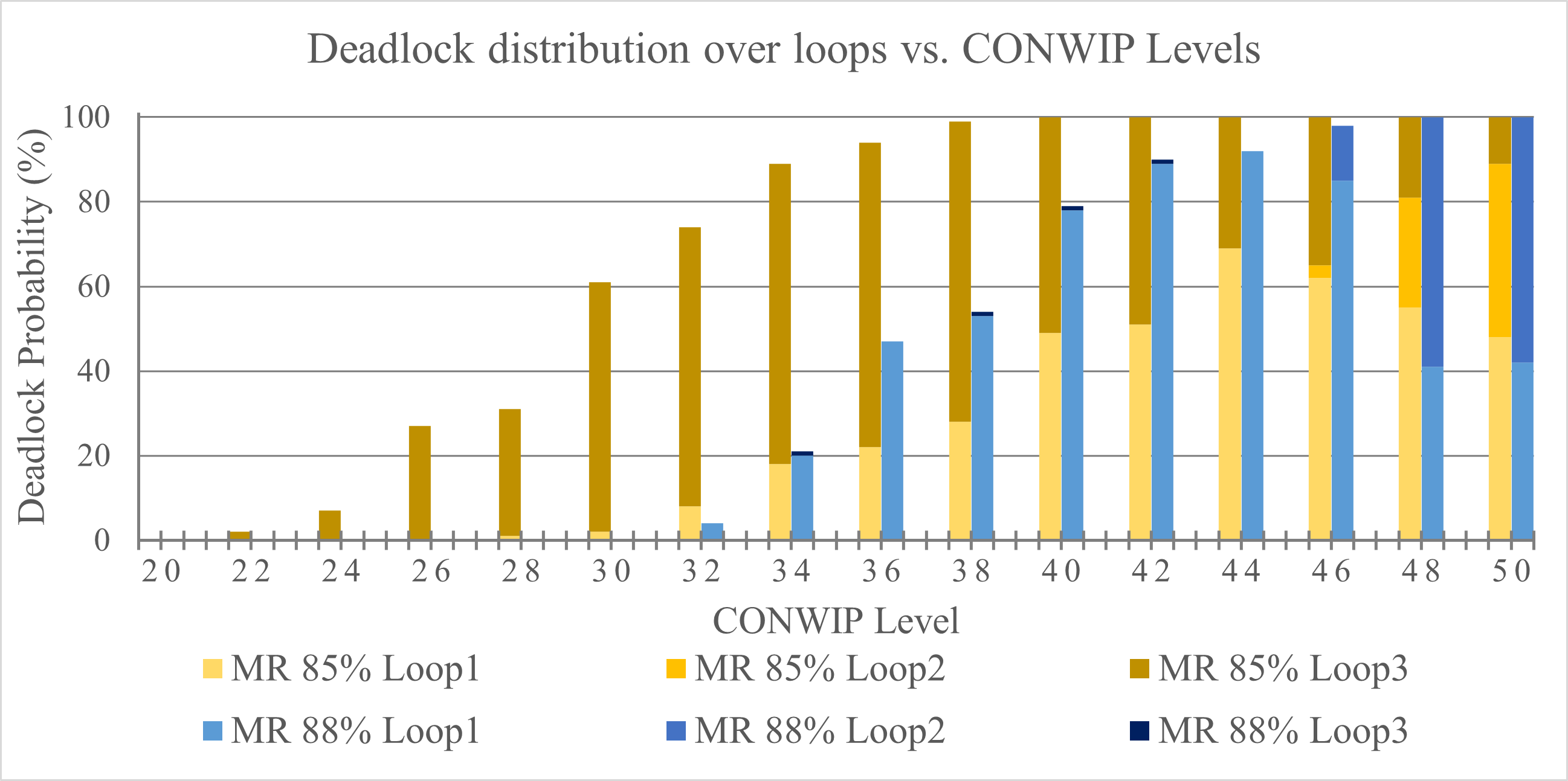}
\caption{Deadlock distribution over the three loops under different MR levels (two sets of stacked bars, for 85\% and 88\%, respectively) and WIP* values. Only scenarios with WSQC=FIFO and SO2=SO3=NO are considered.}
\label{fig:InitialDLLoopImpact}
\end{figure}

Fig. \ref{fig:InitialDLLoopImpact} deepens the analysis of the DL distributions across the three loops (again, only scenarios with QCR=FIFO and SO2=SO3=NO are considered). Different shades of yellow and blue bars show the first loop in which DL occurs in the simulation, respectively for MR=85\% and MR=88\%. The bar height represents the average DL probability in scenarios with a specific WIP*, and, inside each bar, this probability is distributed according to the first loop in which DL occurs. The probability that the first DL occurs in a certain loop changes with WIP*. Such probability also changes with different MR levels. In fact, Fig. \ref{fig:InitialDLLoopImpact} shows that loop 1 is the most critical with MR=88\% (light blue bars), as it is the first probability arising with increasing CONWIP level (WIP* equals to 32). Conversely, with MR=85\%, loop 3 becomes the most critical at lower WIP* (dark yellow bars increase from WIP* = 22). However, with different WIP* value, critical loops shift from one to another (e.g., loop 2 becomes the most critical with high WIP* and MR=88\%). 

\begin{figure}[h!]
    \begin{subfigure}[]{\textwidth}
        \centering
        \includegraphics[draft=false, width=0.8\textwidth]{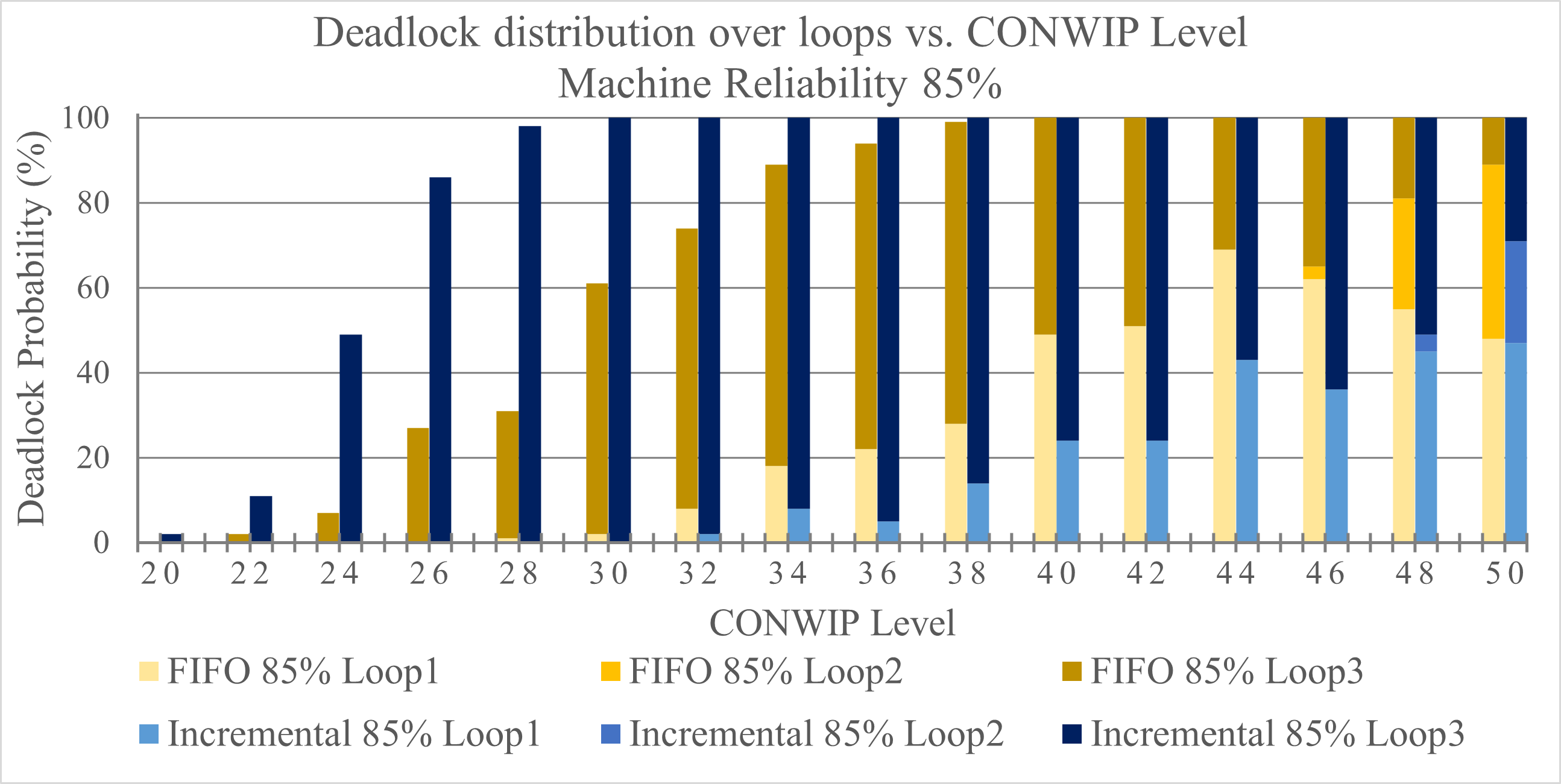}
        \subcaption{MR=85\%.}
        \label{fig:FIFOvsINC85}
    \end{subfigure}
    \begin{subfigure}[]{\textwidth}
        \centering
        \includegraphics[draft=false, width=0.8\textwidth]{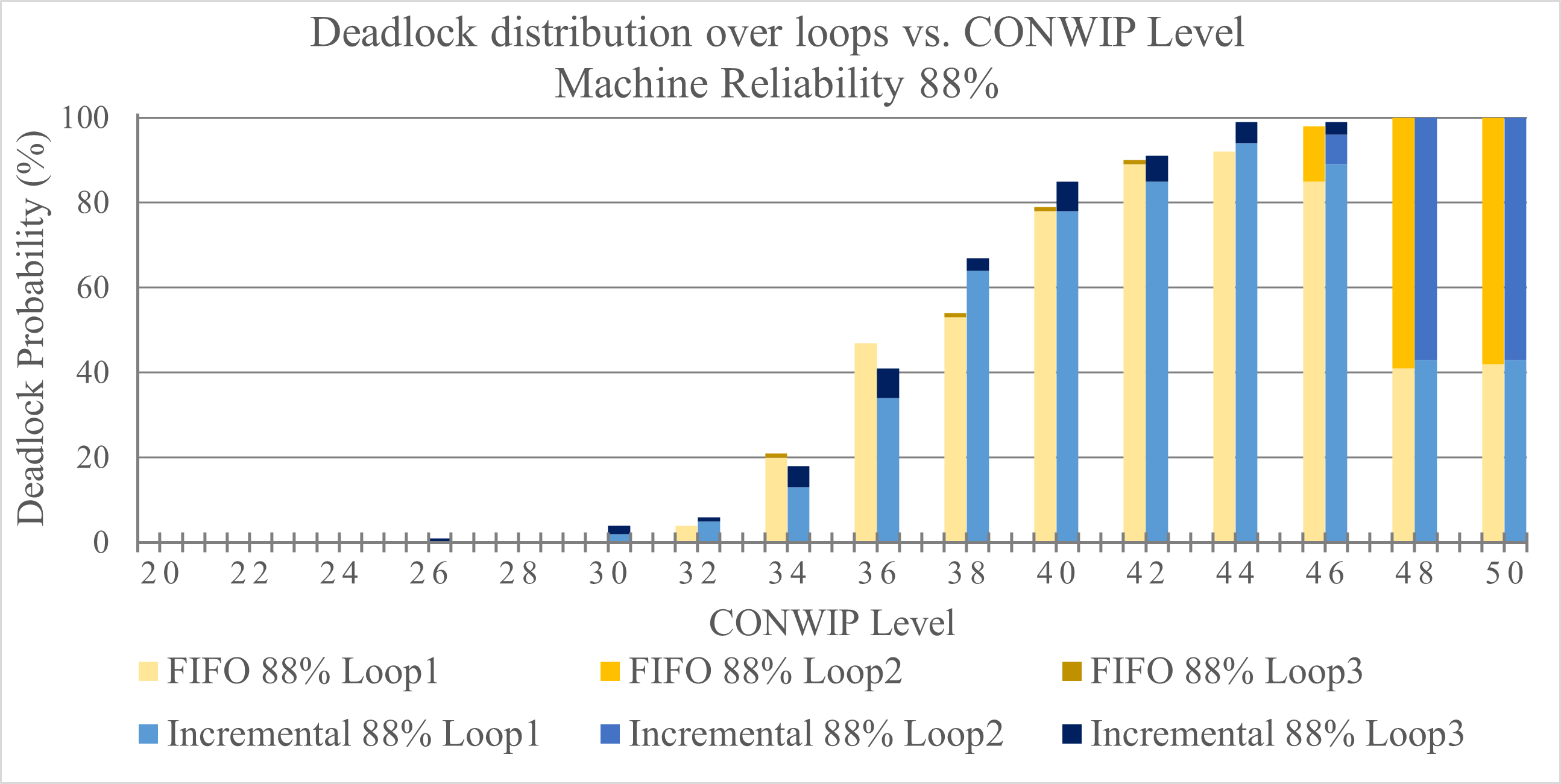} 
        \subcaption{MR=88\%.}
        \label{fig:FIFOvsINC88}
    \end{subfigure}
 \centering
   \caption{Deadlock distribution over the three loops by comparing the impacts on the entire system of \textit{FIFO} and \textit{Inc} policies in the WSQC loop. Only scenarios with SO2=SO3=NO are considered.}
\label{fig:FIFOvsINC}
\end{figure}

Figures \ref{fig:FIFOvsINC} compare the probability that a deadlock occurs in loop 1, loop 2, and loop 3 (with shades of the same colors going from light to dark, respectively) in the case of \textit{FIFO} and \textit{Inc} sequencing rules at WSQC (in yellow and in blue, respectively), and SO2=SO3=NO. Thus, it shows that job sequencing rules provide: (i) a quick proportional reaction to disruptions (in this case, MR deterioration),  (ii) by sharing the overload of a loop towards another one. In particular, Fig. \ref{fig:FIFOvsINC85} shows that, for MR=85\%, the \textit{Inc} rule in WSQC increases the load in loop 3 (dark blue bars higher than dark yellow) by absorbing the overload from loop 1 (light blue bars lower than light yellow ones). The same behavior appears in a smoothed fashion in Fig. \ref{fig:FIFOvsINC88}, in which the variability propagated by the machine reliability is lower (MR=88\%).  

\begin{figure}[ht!]
    \begin{subfigure}[]{\textwidth}
        \centering
        \includegraphics[draft=false, width=0.8\textwidth]{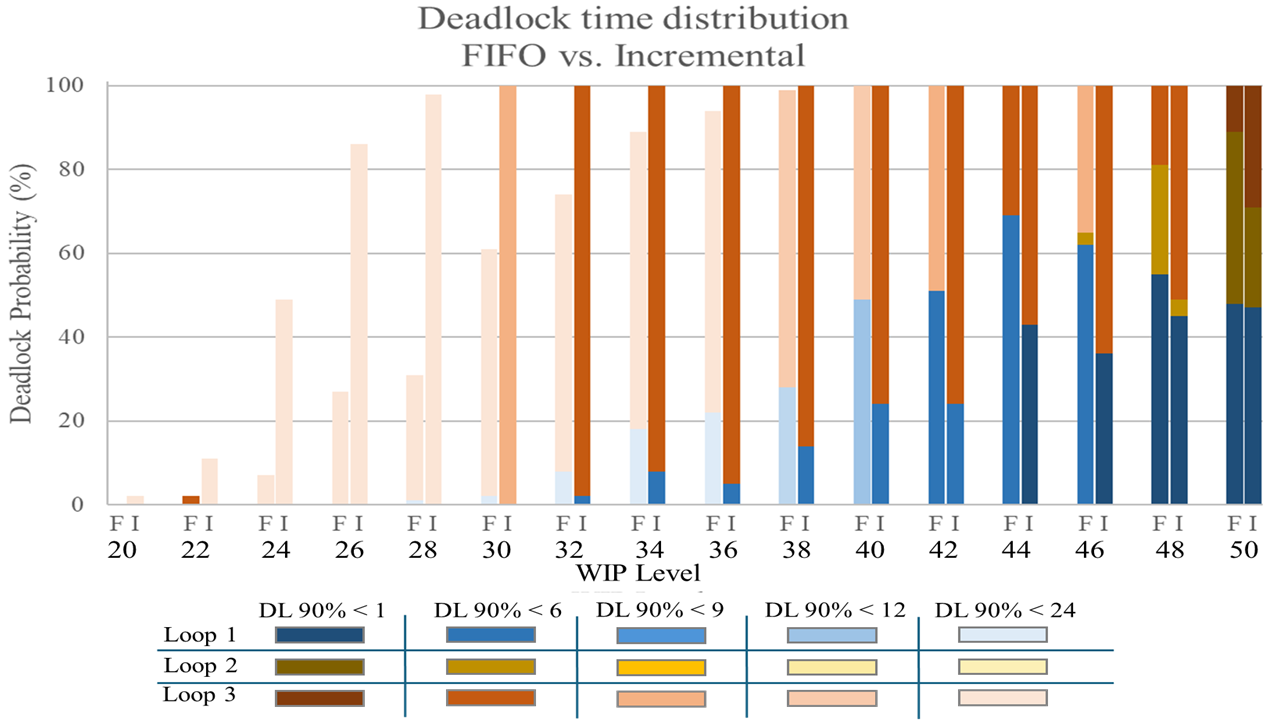}
        \subcaption{MR=85\%.}
        \label{fig:FIFOvsINCtime85}
    \end{subfigure}
    \begin{subfigure}[]{\textwidth}
        \centering
        \includegraphics[draft=false, width=0.8\textwidth]{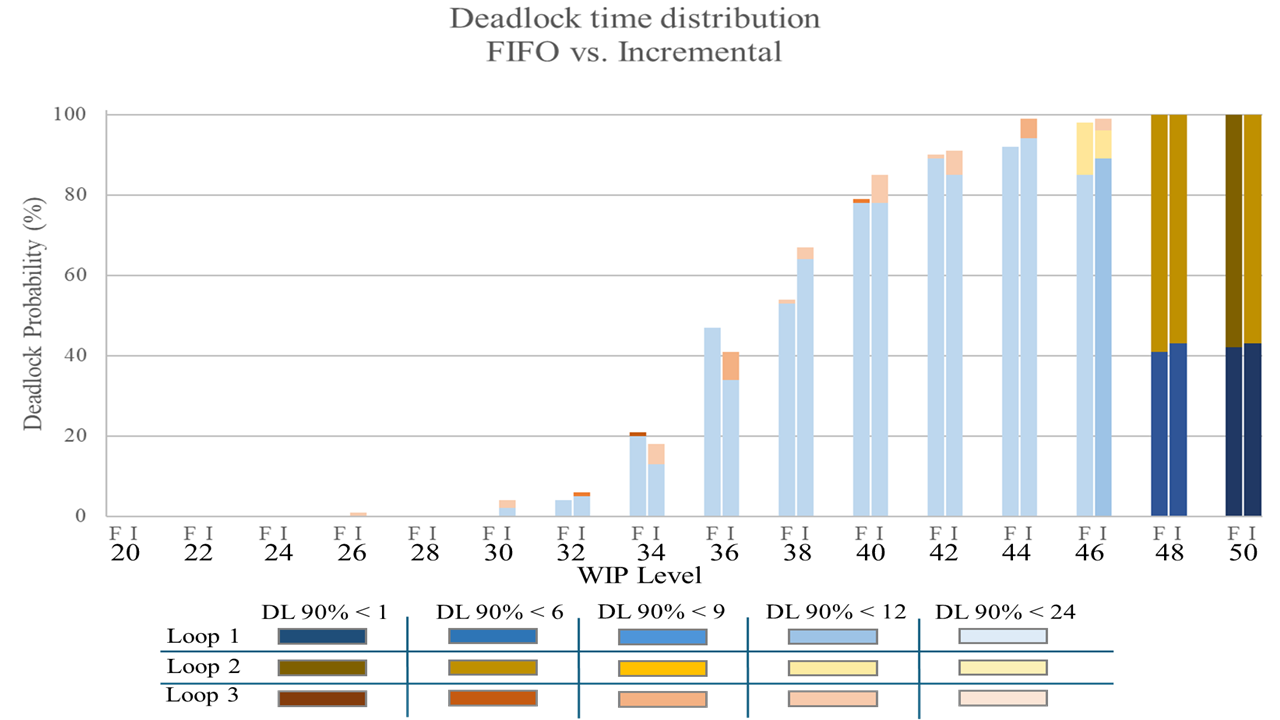} 
        \subcaption{MR=88\%.}
        \label{fig:FIFOvsINCtime88}
    \end{subfigure}
 \centering
   \caption{The impact of \textit{Inc} job sequencing rule (the right bar denoted by "I") in loop 3 on deadlock probability and its distribution over the three loops (blue, yellow, and red, respectively). The color shades represent the severity of deadlocks, from most critical (90\% probability of occurrence within the first month) to least critical (90\% probability of occurrence within the first two years), for each WIP* value.}
\label{fig:FIFOvsInctime}
\end{figure}

While Figures \ref{fig:InitialDLLoopImpact} and \ref{fig:FIFOvsINC}  show which is the critical loop causing the DL occurrence, they do not give any information about its severity. In this paper, it is assumed that the severity of a DL depends on the time of its occurrence. In fact, the earlier a DL occurs, the higher the system load is at that time. A DL in a lightly loaded system leads to high system instability, characterized by effects that unbalance the workload over time, making it more difficult to manage using WIP* values and routing and sequencing policies. To show how fast the workload in a specific loop increases, Fig. \ref{fig:FIFOvsInctime} reports with different colors the time at which DL occurs for scenarios in which SO2=SO3=NO. 

In Fig. \ref{fig:FIFOvsInctime}, for each WIP* value, the two bars show the DL probability for \textit{FIFO} and \textit{Inc} rules (left versus right bars respectively); different colors identify the loop in which DL occurs (loop 1 in blue, loop 2 in yellow, loop 3 in red); last, different shades of the same colors represent whether the DL occurs (with a probability greater than 90\%) before a specific month (e.g., the darkest red color represents scenarios in which DL occurs with a probability greater than 90\% in loop 3 within the first 30 days, i.e., first months, - $DL 90\% < 1$). The simulation always starts with an empty system, as does the production of new products. In such a system, a severe DL occurs before the system reaches the utilization of the typical behavior (i.e., the nominal behavior), as jobs concentrate in a specific loop more quickly than they distribute themselves throughout the system (dark colors in Fig. \ref{fig:FIFOvsInctime}). Instead, a less severe DL occurs when the system works close to the utilization of its typical behavior, when jobs are already spread through the entire system, and only the variability induced by the reliability of the processes leads to concentrating many jobs in a specific part of the system (thus overloading some carousels while the rest remain underutilized).


The comparison between the scenarios with MR equal to 85\% and 88\% (Figures \ref{fig:FIFOvsINCtime85} and \ref{fig:FIFOvsINCtime88}, respectively) shows again that the interdependence between loops 1 and 3 is tight when MR is low, leading to a reciprocal influence in WS performance and job concentration. Furthermore, regarding the timing of DL occurrence, Fig. \ref{fig:FIFOvsINCtime85} shows that the \textit{Inc} sequencing rule quickly and severely congests loop 3 (high probabilities and dark colors) for scenarios with MR=85\%. Conversely, in Fig. \ref{fig:FIFOvsINCtime88} (scenarios with MR=88\%), there is a smoothed job sharing between the two loops that makes loop 1 less critical without suddenly jeopardizing loop 3. In fact, moving from the \textit{FIFO} to the \textit{Inc} rule (left versus right bars for each WIP* value) reduces the overall deadlock probability and the DL probability in loop 1, while moderately increasing the DL probability of loop 3 (all with light color shades).   

\subsubsection{Leveraging control policies to tune the daily throughput}

The following analyses highlight that combining the CONWIP production control rule with job routing and sequencing can improve control over AVG dTH. To this purpose, only scenarios with a probability of deadlock below 10\% are considered, investigating only strategies with acceptable performance. Cases with different MR levels are analyzed separately, as the machine reliability cannot be controlled by any production control policy. However, while maintenance actions (out of scope for this paper) should be taken to increase MR, split management and job-sequencing rules can mitigate the negative effect of low MR on production performance. Two main results related to the combination of the production control rule with both job sequencing and routing are presented in the following: (i) overall improvement of AVG dTH, (ii) flexible control of system performance. The experiments keep the factorial levels (i.e., split management policies, job routing rule, and WIP* value) constant throughout the entire simulation run to investigate their long-term implications. However, in industrial environments, the strategies can be frequently changed to respond to events such as machine failures, demand and input fluctuations, and load balancing. 

Furthermore, several alternative combinations of the factorial levels can lead to the same AVG dTH level. Thus, a visualization priority has been chosen to present the results by directly referring to the minimum production control level that the system must have to achieve a specific AVG dTH. Table \ref{tab:WIP30ex} presents an illustrative example of all the combinations of split management and sequencing rule parameters that achieve different AVG dTH ranges when the system has MR=85\% and WIP*=30. In the table and in the remainder of the section, AVG dTH has been discretized into 0.5-length ranges for readability. 

The table shows that various AVG dTH ranges can be achieved with WIP*=30 (rows). Specifically, without sequencing or split management rules, maintaining such a WIP* value would result in a deadlock probability exceeding 10\%. In fact, the table contains no lines with empty split management rules for both QCR=\textit{FIFO} and QCR=\textit{Inc}. Instead, by tuning the parameters of the proposed policies, the AVG dTH can be varied from 38-38.5 jobs/day to 51-51.5 jobs/day. The combinations of parameter values that allow such an AVG dTH level are shown in the right part of the table. 

\begin{table}[h!]
\centering
\resizebox{\textwidth}{!}{
\begin{tabular}{|c|c|c|c|}
\hline
\textbf{AVG}& \textbf{Color}& \multicolumn{2}{c|}{\textbf{Split management rule (SO2, SO3)}} \\ 
\textbf{dTH range} & \textbf{scheme}& \textbf{QCR = FIFO} & \textbf{QCR = Inc }\\ \hline
38 - 38.5 & \cellcolor{red} &   & (9,3) \\
42.5 - 43&  \cellcolor{red}& & (19,3)\\
43  - 43.5 & \cellcolor{orange} &   & (NO-28-37,3)\\
45.5 - 46 & \cellcolor{red} &  (9,3-7-13); & (9,7-10-13) \\
49  - 49.5 & \cellcolor{red} &   & (19,7)\\
49.5 - 50& \cellcolor{orange} &   & (NO-28-37,7)\\
50 - 50.5& \cellcolor{red}  & (19,3)  & (19,10)\\
50.5 - 51& \cellcolor{orange} & (NO, 3); (19,7-10-13); (28,3); (37,3)  & (NO,10-13); (19,13); (28,10-13); (37,10-13) \\
51  - 51.5 & \cellcolor{cyan} & (NO,7-10-13); (28,7-10-13); (37,7-10-13)& \\
\hline
\end{tabular}}
\caption{All possible combinations of split management and WSQC sequencing rule parameters to achieve specific TH ranges with: WIP*=30, MR= 85\%.}
\label{tab:WIP30ex}
\end{table}

The priority visualization scheme followed in the Table \ref{tab:WIP30ex} and in the remainder of the section, ordered from the highest priority colors to the lowest ones, is:
\begin{itemize}
    \item \textit{Black:} all the scenarios with DL probability equal to or greater than 10\%;
    \item \textit{Green:} a range of AVG dTH reached by just adopting the CONWIP control policy with the indicated WIP* value. 
    \item \textit{Yellow:} a range of AVG dTH reached by just adopting the CONWIP control policy with the indicated WIP* value and the \textit{Inc} job-sequencing rule in WSQC.
    \item \textit{Orange and red:} a range of AVG dTH reached by adopting the CONWIP control policy with the indicated WIP* value, the \textit{Inc} job-sequencing rule in WSQC, split control in loop 3 (orange), or in both loops 2 and 3 (red, less priority than orange).
    \item \textit{Light and dark blue:} a range of AVG dTH reached by adopting the CONWIP control policy with the indicated WIP* value, the \textit{FIFO} job-sequencing rule in WSQC, split control in loop 3 (light blue), or in both loops 2 and 3 (dark blue, less priority than light blue).
    \item \textit{White:} AVG dTH ranges never reached for a given WIP* through any combination of control strategies.
\end{itemize}

\begin{sidewaysfigure}[htp!]
    \includegraphics[width=22cm, height=13cm]{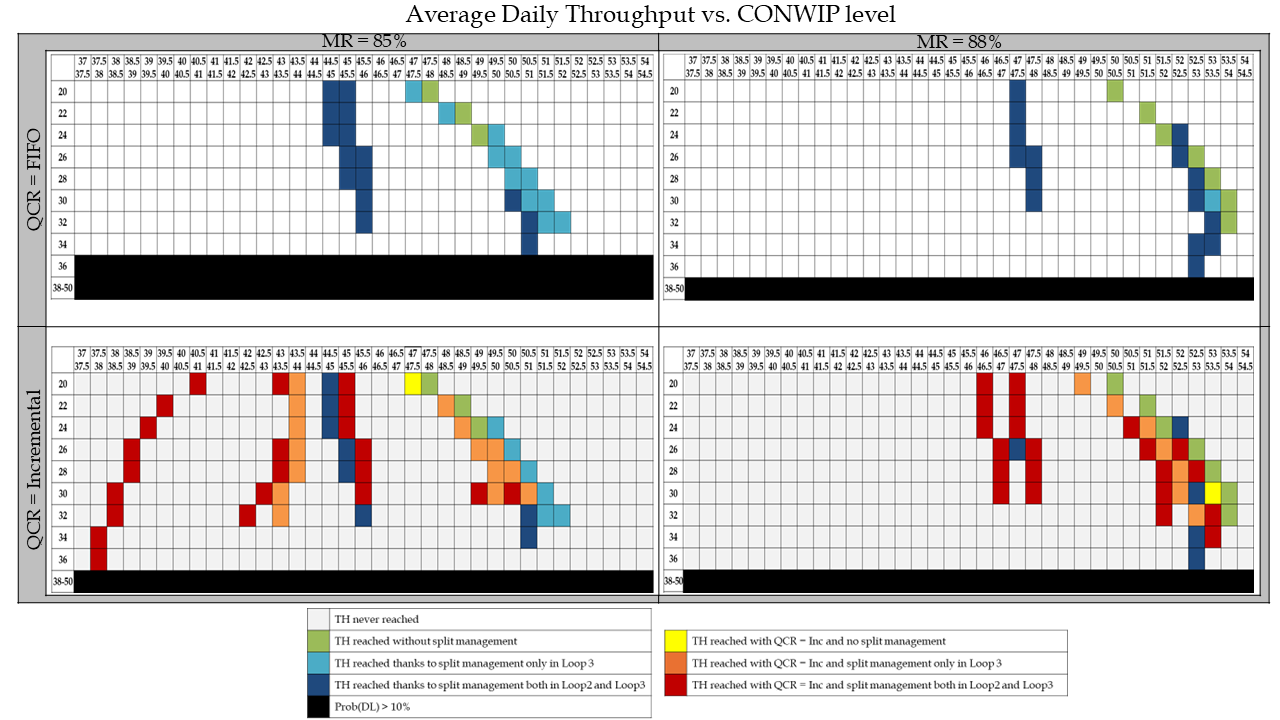}
    \caption{For each WIP* value and for each achievable TH range, the simplest control rules that allow the system to achieve the specific TH range are shown.}
    \label{fig:avgdthvswip}
\end{sidewaysfigure}

The priority color visualization scheme is also exploited in Fig. \ref{fig:avgdthvswip} to show the effect of control policies on AVG dTH for each WIP* value for scenarios with, respectively, MR=85\% and MR=88\%. In particular, Fig. \ref{fig:avgdthvswip} highlights the contribution of the rules for split control by comparing both systems with different machine reliability (comparison of charts in the first column versus the second one) and the additional adoption of job sequencing policies (charts in the first row versus second row). 

The AVG dTH increases with the increase of the WIP* in a pure CONWIP system, up to a WIP* value at which there are no more green squares because the probability of deadlock exceeds 10\%, and, if other strategies are not adopted (i.e., the presence of squares of other colors), the entire line would be black. 

Comparing the charts in column one (MR=85\%) and those in column two (MR=88\%) highlights the impact of machine reliability on the AVG dTH of the pure CONWIP strategy. Less reliable workstations jeopardize the AVG dTH (from 54 to 49.5 jobs/day, i.e., about a 9\% reduction) and the maximum WIP* allowed in the system to keep the probability of deadlock below 10\% (from WIP*=32 to WIP*=24). 

In this context, the first contribution of adopting split control rules (light blue when applied only to loop 3 and dark blue to both loops 2 and 3) is the improvement in the maximum achievable AVG dTH for systems with lower machine reliability. Specifically, in Fig. \ref{fig:avgdthvswip}, the charts in the left-hand column show the AVG dTH increment indicated by the light blue squares. In fact, the minimum AVG dTH with MR=85\% moves from 49.5 to 52 jobs per day, which is much closer to the 54 jobs per day of the system with MR=88\%.

The second contribution of introducing split control rules in the CONWIP production control policy is the enabling of the AVG dTH control while maintaining the fixed WIP* value. Specifically, the adoption of split control rules for loop 2 (dark blue squares) also introduces the capability to control the AVG dTH to react to internal and external changes by acting on job routing and, ultimately, on its flow time. 

Finally, introducing job sequencing rules into WSQC substantially extends the range over which the AVG dTH can be controlled. The ability to control AVG dTH is directly tied to variability affecting the system: the lower the machine reliability (first column, MR=85\%), the more jobs enter loop 3, and the greater the range of AVG dTH control. In fact, in the second column, introducing an incremental sequencing rule only moderately extends the achievable AVG dTH range. High system variability can propagate to both downstream and upstream subsystems, and the interconnected operations, such as inventories and component and raw-material provisioning to the line. Acting on the source of variability (in this system, loop 3, where all defective products enter) reduces the propagation of variability.

\begin{figure}[h!]
    \begin{subfigure}[]{\textwidth}
        \centering
        \includegraphics[draft=false, width=0.9\textwidth]{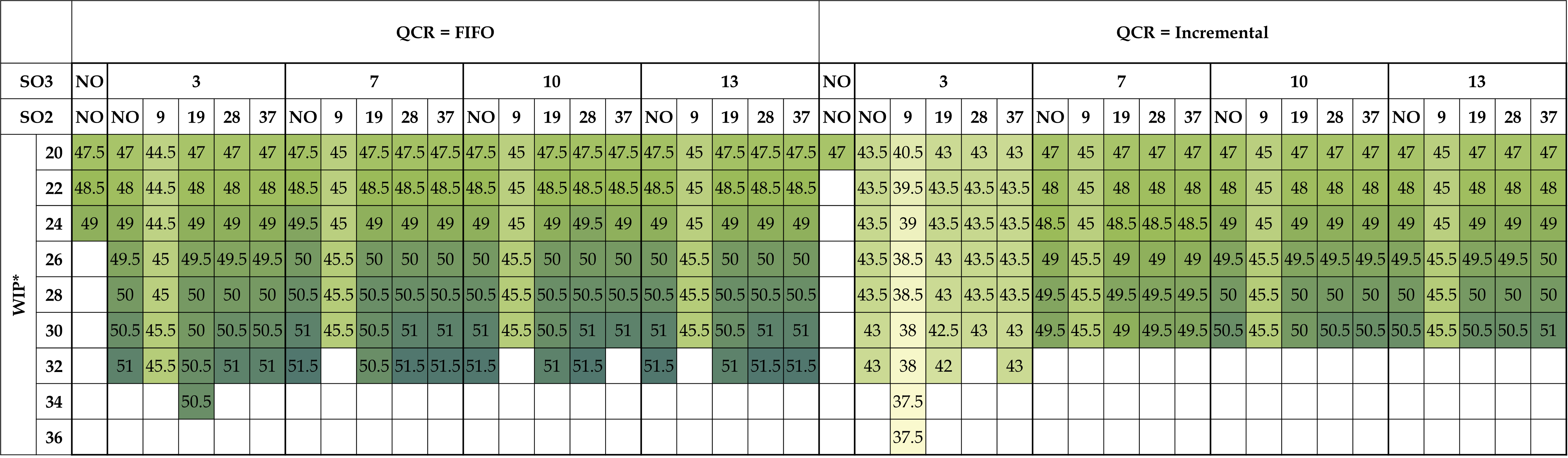} 
        \subcaption{$MR = 85\%$.}\label{fig:split85Param}
    \end{subfigure}
    \begin{subfigure}[]{\textwidth}
        \centering
        \includegraphics[draft=false, width=0.9\textwidth]{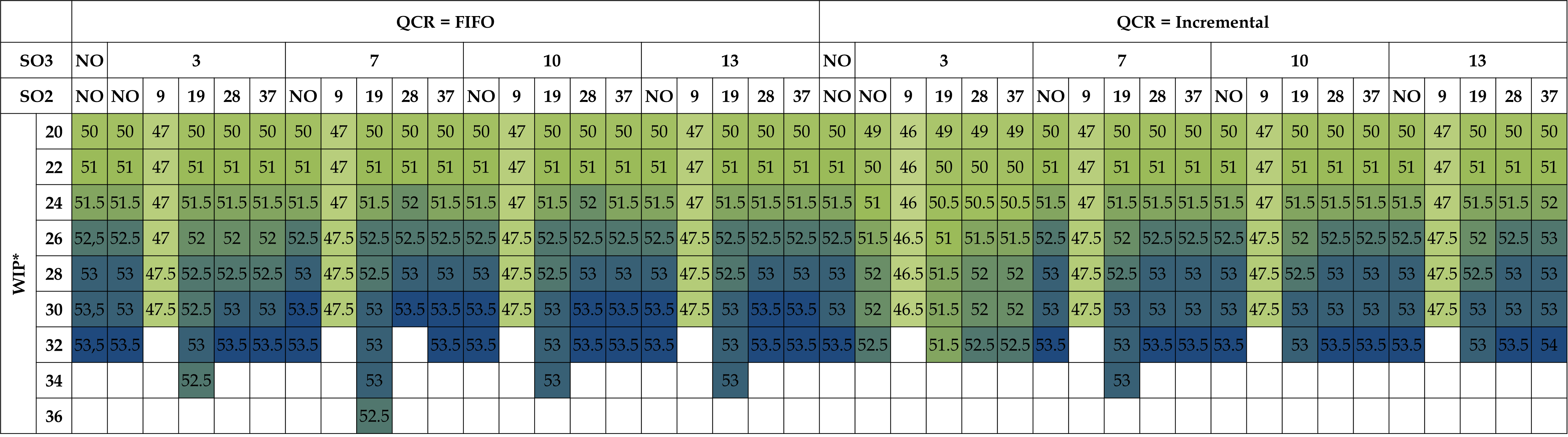}
        \subcaption{$MR = 88\%$.}\label{fig:split88Param}
    \end{subfigure}
   \caption{Heatmap of AVG dTH range for each WIP*, QCR, SO2, and SO3 values. Only scenarios with a deadlock probability less than 10\% are shown.}\label{fig:splitParam}
\end{figure}

More in depth, Fig. \ref{fig:splitParam} shows the AVG dTH achievable by setting the value of SO2 and SO3 parameters (only for scenarios with deadlock probability less than 10\%). Specifically, the two heat maps show the AVG dTH ranges for each WIP* value (in rows) and for each combination of QCR, SO2, and SO3 values (in columns), respectively for MR=85\% (Fig. \ref{fig:split85Param}) and MR=88\% (Fig. \ref{fig:split88Param}). 

Fig. \ref{fig:split85Param} shows that controlling SO3 is important for improving WIP* value while limiting deadlock probability at 10\%. In fact, the first column, with both SO2 and SO3 equal to "NO", can have at most WIP*=24, corresponding to 49 jobs per day. Controlling SO3 to limit the number of jobs in loop 3 allows for an increase in WIP*, leading to higher AVG dTH. Controlling the SO2 maximum number of jobs admissible in loop 2 improves AVG dTH with the increase of SO2, only at high system saturation levels (WIP*$\geq$30), except for the tight limitation level of SO2=9. In fact, when loop 2 is limited (i.e., S2=9), the entire AVG dTH is reduced. Similarly, the job sequencing rule extends the control of AVG dTH, particularly when MR=85\%, by increasing the opportunity to reduce AVG dTH without changing WIP*.Together with an AVG dTH reduction, the average flow time increases (as the Little Law suggests). Moreover, the effect of SO2 tuning appears to be more pronounced than that of SO3: the heat map shows very similar color patterns across all SO3 values.   

\section{Discussion}
\label{sec:discussion}

Sequencing and scheduling play a pivotal role in the context of the CONWIP control policy, particularly in make-to-order environments and, more generally, in all those cases in which jobs are not perfectly identical. However, a further benefit investigated in this paper, and applicable to systems with identical jobs, is the control of workload balancing over time. Controlling the workload distribution across the system is particularly critical in the investigated line due to the presence of deadlocks, and backlog sequencing and scheduling are ineffective because the jobs are identical in the load bay. Conversely, the results show that sequencing jobs already into the system can effectively control the workload distribution across workstations and, when coupled with job routing, align actual throughput with current production targets. Therefore, the results highlight that job sequencing, beyond the most common backlog sequencing and scheduling, is a means for reacting to the dynamic, and sometimes progressive, unbalancing of the workload.

The progressive workload unbalancing caused by the sources of variability also affects the system performance. Specifically, the results show that, given the system configuration, an optimal WIP* value exists (Fig. \ref{fig:avgdthvswip} shows that the maximum AVG dTH is reached at WIP*=32) to maximize AVG dTH. However, additional sources of variability, such as deterioration in MR, make it difficult to reach the optimal WIP* because limited buffer capacity (i.e., space in the three conveyor carousels) leads to deadlocks. Job routing counterbalances the negative effects of the progressive workload unbalancing, making it possible to reach the optimal WIP* that maximizes AVG dTH, thereby reducing the loss in AVG dTH when compared to the system with a higher MR. The rigid structure of the job-handling system, namely the conveyor carousels, makes the line particularly susceptible to progressive workload imbalance arising from asymmetric processing time distributions. In fact, frequent small deviations from the average processing time generate cumulative effects that are difficult to counterbalance without external control mechanisms \citep{Alfieri:2024}.   

The combined use of job sequencing and job routing enables control of the AVG dTH to limit the propagation of variability across upstream and downstream linked subsystems, serving as a decoupling point without changing the number of CONWIP cards. In fact, the number of cards is often represented by complex entities such as pallets or entire trucks \citep{Silva:2015}. Moreover, the need to slow throughput can be limited to a short time window, for example, to allow a quick check at a workstation or to allow a momentary stop by an operator. Job sequencing and job routing can have a quasi-immediate impact on the TH. Conversely, changing the number of cards can introduce an unpredictable lag between the action and the effective reduction in TH, and may also require a longer undefined lag to return the system to its initial target productivity. 

The experimental results show that job routing and job sequencing allow production control in a CONWIP-driven environment without necessarily changing the number of cards. However, leveraging job routing and sequencing does not always lead to predictable results, as the latter depend on the initial system state in which they are activated. Finally, this paper investigates the use of a single combination of job sequencing and routing, showing its effectiveness in controlling production. However, when varying the control policy over time, the magnitude of the effects and the lag before they occur are not predictable, and, in the worst case, the policy change itself can be counterproductive.   

\subsection{Managerial Insights}

The experimental results highlight that dynamic CONWIP should not focus solely on reacting to or adapting to external changes, such as demand fluctuations, but also on the progressive unbalancing of the system workload. In the system design phase, the system should be tested under all potential causes and locations of workload imbalance to properly design buffers, raw material supplies to the workstations, and overall operational coordination to address them.

From a managerial point of view, leveraging job sequencing and routing to control system performance through workload balancing under a CONWIP policy requires the proper identification of KPIs, the ability to monitor them in quasi-real time, and the ability to physically activate the specific control rules. The experimental results indirectly shed light on the risk of implementing the CONWIP policy without a digital infrastructure that enables proper system monitoring and the ability to take actions to control job flows. Moreover, leveraging approximate system models and proxy KPIs, such as the TH at the system exit point, can lead to control mechanisms that are unable to properly detect progressive workload unbalancing and react promptly.

Among the main system aspects that should be monitored to properly control production through job sequencing and routing for a dynamic CONWIP, a pivotal role is the intersection between job-handling and manufacturing systems. This paper investigated a continuous-flow, fixed-path handling system, but the managerial insights can be extended to other job-handling systems, buffer sizes and their positioning, and the use of transport batches. In fact, Fig. \ref{fig:benchmark} clearly shows the negative effects of implementing conveyor carousels rather than buffered workstations, and more generally, the impact of job-handling systems. Job-handling systems should be chosen by also evaluating their ability to support production control by creating protective WIP when and where necessary, while enabling the opportunity of redistributing workload while keeping a fixed number of cards.

Finally, discrete event simulation plays a pivotal role also beyond the design phase of both the control mechanisms and the physical system itself. At the operational level, simulation is important for properly evaluating and deciding when and how to change the control policy (e.g., different combinations of job sequencing and routing). In fact, the effectiveness of any production control policy depends critically on the initial state to which it is applied, and simulation enables evaluation of the potential outcomes of each policy in advance.

\section{Concluding Remarks}
\label{sec:conclusion}
This paper investigates the integrated use of a pure CONWIP production control policy in combination with job sequencing and job routing, shedding light on the nature of dynamic production control under CONWIP control. Specifically, dynamic production control using the CONWIP approach is a means of workload balancing in which external changes affect system efficacy, while internal variability undermines effectiveness over time. In particular, the investigated system employs conveyor carousels to move jobs among the workstations, serving as a sort of circulating buffer. 

The paper shows that the employed job-handling system can deeply affect the system performance. In particular, machine reliability (and other sources of variability) can progressively unbalance the workload distribution among conveyor carousels, increasing congestion up to the deadlock state (i.e., a condition in which the carousel is full, preventing the workstations from unloading the processed job, and requiring the system to stop and the intervention of a maintenance team to restore the line operation). Deterioration in machine reliability increases system variability, and the combined effect with conveyor carousels further degrades system performance under a pure CONWIP policy. In fact, workload unbalancing prevents the system from using the optimal WIP level that maximizes performance while avoiding deadlocks.

Sequencing jobs already in the system (rather than the backlog) and implementing job-routing control can effectively counterbalance the progressive workload imbalance without changing the number of cards, which can sometimes be difficult or impractical. 

The experimental results indirectly highlight that dynamic CONWIP approaches require a digital infrastructure for quasi-real-time monitoring and then production control. Missing infrastructure or approximate approaches may fail to detect workload imbalance and can adversely affect system performance. Moreover, discrete-event simulation is pivotal in operational activities, where it is difficult to predict the outcome of changing production policy (e.g., a combination of job sequencing and routing) because it depends on the current system state.

This paper contributes to the scientific literature by investigating the impact of interdependence between the job-handling and manufacturing systems on the CONWIP control policy. It aims to shed light on the progressive workload imbalance across the system, providing empirical evidence for scholars and practitioners interested in dynamic production control. Moreover, the paper provides managerial insights for designing systems that leverage job sequencing and routing to align system performance with targets, while limiting the propagation of variability to interconnected subsystems and avoiding nervousness caused by frequent card changes.

The contribution of this paper can be extended to most of the methods of dynamic production control. However, systems operating in a make-to-order paradigm, or more generally, in an environment where jobs are not identical, require further investigation. In fact, this paper does not deepen any aspect of job flow time or other individual job KPIs, such as lateness.

Future research will investigate the local WIP control among the loops. In particular, hybrid production control mechanisms will be studied by combining CONWIP and Kanban strategies to separately manage the overall WIP entering the system (global approach) and the WIP distribution between loops (local approach). Moreover, further research will investigate the control mechanism to dynamically assess whether varying the production policy, or actually changing the control policy itself, leads to a lag between the policy change and its effects.

    \section*{Acknowledgement(s)}
This study was carried out within the Motown: Smart Production Planning and Control for Manufacturing of Electric Vehicle Powertrain in Industry 4.0 Environment project – funded by European Union – Next Generation EU, Missione 4 Componente 1 CUP E53D23007900006, within the PRIN 2022 program (D.D. 104 - 02/02/2022 Ministero dell’Università e della Ricerca). This manuscript reflects only the authors’ views and opinions and the Ministry cannot be considered responsible for them.

\section*{Disclosure statement}
The authors report there are no competing interests to declare.

 \bibliographystyle{tfcad} 
 \bibliography{bibliography}

\end{document}